# Stable aqueous dispersions of optically and electronically active phosphorene

*Joohoon Kang[1], Spencer A. Wells[1], Joshua D. Wood[1], Jae-Hyeok Lee[1], Xiaolong Liu[2], Christopher R. Ryder[1], Jian Zhu[1], Jeffrey R. Guest[3], Chad A. Husko[3], and Mark C. Hersam[1,2,4,5,6]\**

[1]Department of Materials Science and Engineering, Northwestern University, Evanston, IL 60208, USA

[2]Graduate Program in Applied Physics, Northwestern University, Evanston, IL 60208, USA

[3]Center for Nanoscale Materials, Argonne National Laboratory, Argonne, IL 60439, USA

[4]Department of Chemistry, Northwestern University, Evanston, IL 60208, USA

[5]Department of Medicine, Northwestern University, Evanston, IL 60208, USA

[6]Department of Electrical Engineering and Computer Science, Northwestern University, Evanston, IL 60208, USA

[*]Correspondence should be addressed to: m-hersam@northwestern.edu.

**Abstract**

Understanding and exploiting the remarkable optical and electronic properties of phosphorene require mass production methods that avoid chemical degradation. While solution-based strategies have been developed for scalable exfoliation of black phosphorus, these techniques have thus far employed anhydrous organic solvents in an effort to minimize exposure to known oxidants, but at the cost of limited exfoliation yield and flake size distribution. Here, we present an alternative phosphorene production method based on surfactant-assisted exfoliation and post-processing of black phosphorus in deoxygenated water. From comprehensive microscopic and spectroscopic analysis, this approach is shown to yield phosphorene dispersions that are stable, highly concentrated, and comparable to micromechanically exfoliated phosphorene in structure and chemistry. Due to the high exfoliation efficiency of this process, the resulting phosphorene flakes are thinner than anhydrous organic solvent dispersions, thus allowing the observation of layer-dependent photoluminescence down to the monolayer limit. Furthermore, to demonstrate preservation of electronic properties following solution processing, the aqueous-exfoliated phosphorene flakes are employed in field-effect transistors with high drive currents and current modulation ratios. Overall, this method enables the isolation and mass production of few-layer phosphorene, which will accelerate ongoing efforts to realize a diverse range of phosphorene-based applications.

**Significance Statement**

Few-layered phosphorene, which is isolated through exfoliation from black phosphorus, has attracted great interest due to its unique electronic and optical properties. Although solution-based exfoliation methods have been developed for black phosphorus, these techniques have thus far employed anhydrous organic solvents. This approach minimizes exposure to known oxidizing species, but at the cost of limited exfoliation yield and relatively thick flakes. Here, we overcome these limitations by employing stabilizing surfactants in deoxygenated water, which results in phosphorene down to the monolayer limit. The resulting aqueous phosphorene dispersions show layer-dependent photoluminescence, and enable high performance field-effect transistors. Overall, this approach holds promise for the solution-phase production of few-layered phosphorene in emerging large-volume applications including electronics and optoelectronics.

\body

Few-layer phosphorene (FL-P) isolated by micromechanical exfoliation has been widely studied both fundamentally and in applications such as high-performance electronic and optoelectronic devices (1-11). Although micromechanical exfoliation provides individual, high-quality FL-P flakes, this technique lacks scalability and is not amenable to large-area applications. Conventional approaches for mass production of two-dimensional (2D) nanomaterials involve chemical vapor deposition (CVD) and liquid phase exfoliation (LPE). Whereas CVD growth of black phosphorus (BP) thin films is hindered by challenges with molecular precursors and extreme growth conditions (12), LPE of BP has been demonstrated and used for the large-scale deposition of thin films akin to the approaches for other 2D nanomaterials (13-17). Specifically, stable BP dispersions have been produced by LPE using high boiling point solvents including *N*-methyl-2-pyrrolidone (NMP), dimethylformamide, dimethyl sulfoxide, and *N*-cyclohexyl-2-pyrrolidone (18-21). With these anhydrous organic solvents, chemical degradation from ambient $O_2$ and water (11) are avoided, but the exfoliation yield and flake size distribution are suboptimal, especially compared to the results that have been achieved with other 2D nanomaterials using stabilizing surfactants in aqueous solutions (16). Furthermore, organic solvents have limited compatibility with methods such as ultracentrifugation for structural fine tuning and sorting, and their high boiling points and safety issues present post-processing challenges (13, 15).

Here, we establish a scalable, high-yield, and environmentally benign method for preparing FL-P via ultrasonication in deoxygenated water. Fig. 1*A* depicts the procedure for obtaining FL-P enriched dispersions. To minimize BP chemical degradation, deoxygenated water is prepared by purging deionized water with ultrahigh purity Ar gas in a sealed container (18, 22). Bulk BP crystals (SI Appendix, methods) are then exfoliated with tip ultrasonication in a vessel sealed in an Ar environment that contains deoxygenated water with 2% w $v^{-1}$ sodium dodecylsulfate (SDS). For FL-P enrichment, as-prepared BP solutions undergo two steps of sedimentation-based ultracentrifugation to isolate thin flakes with relatively large lateral area (SI Appendix, methods). Initially, the as-exfoliated solution is dark brown (Fig. 1*B*, left), softening to light pale yellow after solution dilution and thick flake removal (Fig. 1*B*, right).

The chemical, structural, and optical properties of the resulting FL-P solutions are ascertained through transmission electron microscopy (TEM), Raman spectroscopy, photoluminescence (PL) spectroscopy, and X-ray photoelectron spectroscopy (XPS). Fig. 1*C* shows a representative, low-magnification TEM image of a FL-P nanosheet, while the high-resolution TEM (HRTEM) image of Fig. 1*D* reveals the atomic structure. Selected area electron diffraction (SAED) patterns (Fig. 1*E*) confirm that the FL-P nanosheets are crystalline and orthorhombic in structure. The HRTEM and SAED data further show no evidence of structural disorder or oxidation. In Fig. 1*F*, a liquid-phase Raman spectrum from the FL-P dispersion shows the three representative BP Raman modes at ~362 cm$^{-1}$ ($A_g^1$), ~439 cm$^{-1}$ ($B_{2g}$), and ~466 cm$^{-1}$ ($A_g^2$), with corresponding full-width at half-maximum (FWHM) values of 3.0, 3.3, and 5.1 cm$^{-1}$, respectively (4, 7). Monolayer and few-layer phosphorene flakes have higher FWHM values than thicker (>5 nm) BP (4), making the FWHM broadening compared to precipitated, thicker BP dispersions indicative of FL-P enrichment in our process (SI Appendix, Fig. S1) (4). A visible (Si CCD) PL spectrum of the FL-P dispersion taken with 532 nm excitation reveals an emission peak at ~1.37 eV, consistent with previous reports for monolayer phosphorene (Fig. 1*G*) (7). Finally, the chemical quality of the FL-P flakes are assessed using XPS in Fig. 1*H*. XPS shows that FL-P exhibits the P 2p$^{3/2}$ and P 2p$^{1/2}$ doublet, characteristic of crystalline BP. Weak oxidized phosphorus (i.e., PO$_x$) sub-bands are also observed at ~136 eV in agreement with previous measurements of electronic-grade BP (11, 18, 23, 24). XPS taken on FL-P prepared with other conventional surfactants (e.g., Pluronic F68 and sodium cholate) display similar results (SI Appendix, Fig. S2).

Although BP has been described as highly hydrophilic (25), the FL-P dispersion is only stabilized with the incorporation of amphiphilic surfactants in aqueous solution, similar to other hydrophobic 2D nanomaterials (13-17). BP dispersions prepared with SDS and without SDS were sealed with Ar and left to settle overnight. While an aqueous BP dispersion prepared with SDS results in a stable dark brown solution (Fig. 2*A*, left), BP dispersed in water without surfactants precipitates quickly (Fig. 2*A*, right). To clarify this apparent contradiction, the hydrophilicity of the BP surface was measured with contact angle measurements on a freshly cleaved flat BP crystal. Fig. 2*B* shows that, immediately following exfoliation, the BP surface has an average contact angle of ~57°, indicating that the hydrophilicity of BP is between graphene oxide (~27°) and other 2D nanomaterials (~90°) such as graphene and transition metal dichalcogenides (14, 26). From the dispersions in Fig. 2A, the supernatants were carefully decanted and utilized for optical absorbance and zeta potential measurements. From the higher optical absorbance (Fig. 2*C*) and lower zeta potential value (Fig. 2*D*), the relatively hydrophobic, freshly exfoliated BP nanosheets are stabilized in aqueous solution with amphiphilic surfactants in a manner analogous to other 2D nanomaterials.

Most 2D nanomaterials show high exfoliation yield and stability when dispersed with surfactants in aqueous solution compared to surfactant-free organic solvents (13, 15). To investigate this stability trend with BP, LPE in both SDS-water and NMP was performed under identical exfoliation and centrifugation conditions. Centrifugation in each case occurred in steps ranging from 500 r.p.m. to 15,000 r.p.m. in order to compare the concentration of the resulting dispersions. Fig. 2*E* and 2*F* show that the BP dispersion in NMP possesses a lighter yellow color, indicative of a lower concentration, compared to BP in SDS-water. The actual concentration of BP in SDS-water was calculated from the optical absorbance at 660 nm using the measured extinction coefficient shown in the SI Appendix, Fig. S3 (the extinction coefficient that was used for BP in NMP was reported in ref. 6). This analysis concluded that the concentration of BP in SDS-water is approximately an order of magnitude higher than that of BP in NMP after centrifuging at 15,000 r.p.m. (Fig. 2*G*) (18). Atomic force microscopy (AFM) images of BP exfoliated in SDS-water (red) and NMP (green) followed by 7,500 r.p.m. centrifugation are shown in Fig. 2*H* and 2*I*. The thickness histogram in Fig. 2*J* reveals that FL-P nanosheets prepared in SDS-water have a tighter thickness distribution and thinner average thickness (4.5 nm compared to 17.6 nm) than BP prepared in NMP. Additional thickness and area histograms for FL-P prepared in SDS-water with smaller bins are provided in the SI Appendix, Fig. S4. These results illustrate the effectiveness of aqueous surfactant solutions for producing thin FL-P nanosheets compared to organic solvents.

Following FL-P solution preparation, optical absorbance spectra were measured (Fig. 3*A*). A higher resolution optical absorbance spectrum shows two peaks between 900 nm and 1250 nm (Fig. 3*A*, inset), which is consistent with previous reports for monolayer and bilayer phosphorene (19). PL spectroscopy was also performed on the FL-P solution using an excitation wavelength of 532 nm and a Si CCD for emission wavelengths up to 1000 nm and a $N_2$-cooled InGaAs array for emission wavelengths between 1000 nm and 1600 nm. In Fig. 3*B*, the measured PL spectra were fit with Gaussian functions at peak positions of ~907 nm (red area), ~1215 nm (orange area), and ~1428 nm (green area) (7, 27). To correlate these peak positions with the number of phosphorene layers, micromechanically exfoliated BP flakes were prepared on a 300 nm $SiO_2$/Si substrate and then passivated with thin $Al_2O_3$ films (11). PL spectra and corresponding optical microscopy images for flakes ranging in thickness between one layer (1L) and five layers (5L) are shown in the SI Appendix, Fig. S5 to S8. Based on the PL peak positions measured for the micromechanically exfoliated BP flakes, the peaks at ~907 nm, ~1215 nm, and ~1428 nm from the FL-P dispersion (Fig. 3*B*) are assigned to 1L, 2L, and 3L+ phosphorene, respectively.

Additionally, solid-state PL spectra were measured on FL-P aggregates deposited onto a $SiO_2$/Si substrate using an excitation wavelength of 532 nm. The measured PL spectra have a 1L peak position at ~909 nm, as shown in the SI Appendix, Fig. S9*A*. In Fig. S10, additional peaks were observed for 2L at ~1261 nm and 3L+ at ~1459 nm, ~1528 nm, and ~1588 nm using a thermoelectrically cooled InGaAs array. The positions of the PL peaks from FL-P are indicated in the SI Appendix, Table S1. The peak position differences between the liquid-phase sample and the solid-state sample can be attributed to the different dielectric screening from the surrounding environment (7, 28).

To verify that the observed PL is not defect-mediated (29, 30), FL-P aggregates were exposed to UV ozone to intentionally introduce oxygen defects. As the UV ozone exposure time increases, the PL emission intensity at ~909 nm decreases and a $PO_x$ defect-mediated emission peak emerges at ~780 nm (SI Appendix, Fig. S9*D*). The emission intensities at ~909 nm (red curve) and ~780 nm (green curve) are plotted as a function of UV ozone exposure time, with the 909 nm peak monotonically decreasing as the FL-P is oxidized, and the 780 nm peak initially increasing in intensity due to increased defect concentration and ultimately decaying due to the complete destruction of the material (SI Appendix, Fig. S9*E*). This observation corresponds closely to the time-evolution of PL with respect to photo-oxidation time for micromechanically exfoliated BP (SI Appendix, Fig. S5*D*). Furthermore, PL spectra were measured under the same excitation conditions on the FL-P dispersion before (red area) and after oxidation introduced by $O_2$ gas bubbling (blue area). Following oxidation, the FL-P shows significantly decreases in PL intensity (Fig. 3*C*, left) in addition to substantial increases in the $PO_x$ XPS peak (Fig. 3*C*, right). Consequently, the PL peaks observed in Fig. 3*B* can be attributed to 1L, 2L, and 3L+ phosphorene as opposed to defects introduced by solution processing.

To explore the electrical properties of individual FL-P nanosheets, field-effect transistors (FETs) were fabricated by electron-beam lithography (EBL). Prior to FET fabrication, lateral size sorting of the FL-P solution was performed to enrich large FL-P nanosheets that can bridge the 200 nm electrode gap. Lateral size sorting of the FL-P solution was achieved by sedimentation-based density gradient ultracentrifugation (s-DGU). 3 mL of as-prepared FL-P dispersion was carefully placed on top of a 10 mL linear density gradient (1.05 to 1.10 g cm$^{-3}$) formed using deoxygenated iodixanol that was loaded with 2% w v$^{-1}$ SDS. Ultracentrifugation was then performed at 10,000 r.p.m. for 1 hr at 22 ºC using a SW41 Ti rotor (Beckman-Coulter). Following ultracentrifugation, fractionation was achieved using a piston gradient fractionator (BioComp Instruments). By using s-DGU, the average flake area was tuned by over an order of magnitude (Fig. 4*A*). Each fraction (f1 to f6 from the top to the bottom of the centrifuge tube) was collected and deposited onto Si substrates for AFM measurements. The AFM images of f1, f3, and f5 clearly show the size differences between each fraction (SI Appendix, Fig. S11). From the AFM measurements, the average areas of f3 to f6 were found to be appropriate for FET fabrication (Fig. 4*B*).

To prepare arrays of the FL-P nanosheets, fractions f3 to f6 were collected on anodic aluminum oxide (AAO) membranes by vacuum filtration, rinsed with deoxygenated water to remove excess surfactant (SI Appendix, Figs. S12 and S13 for AFM and XPS measurements before and after deoxygenated water rinsing), and transferred onto degenerately doped Si substrates coated with 20 nm of atomic layer deposition (ALD) $Al_2O_3$ or $HfO_2$. Following the FL-P transfer, Au/Ni electrodes of 10 μm width and 200 nm length were patterned using EBL. A false-colored scanning electron microscopy (SEM) image (Fig. 4*C*) shows several FL-P nanosheets (green) connected in parallel between two electrodes (yellow). FET output and transfer curves (Fig. 4*D* and 4*E*) reveal ambipolar behavior for FL-P with a current modulation ($I_{ON}/I_{OFF}$ ratio) of ~5 x 10$^3$ and maximum drive current ($I_{DS}$) of ~130 µA µm$^{-1}$ at $V_{DS}$ = 1 V (the transfer curve for the forward and reverse gate voltage sweep is shown in the SI Appendix, Fig. S14). Histograms of the maximum drive current and current modulation from several FL-P FETs are shown in the SI Appendix, Fig. S15. For comparison, $I_{ON}/I_{OFF}$ ratio and maximum $I_{DS}$ values for BP FETs from the recent literature (both micromechanically exfoliated and organic solvent exfoliated) (2, 5, 11, 18, 31-33) are plotted in Fig. 4*F*. Results from the FL-P device from Fig. 4*E* are also shown as a yellow star on this plot. Since the upper right hand corner of this plot represents an optimal combined performance from the perspective of maximum drive current and current modulation, it is evident that aqueous-exfoliated FL-P compares favorably with competing BP exfoliation methods from the perspectives of electronic properties and utility for electronic devices.

In summary, effective exfoliation of FL-P nanosheets has been achieved by ultrasonication in deoxygenated water stabilized with surfactants. This method results in stable, highly concentrated few-layer phosphorene with distinct advantages compared to exfoliation in organic solvents. Comprehensive microscopic and spectroscopic analysis shows that individual FL-P nanosheets possess properties comparable to micromechanically exfoliated BP flakes without chemical degradation following aqueous processing. For example, PL measurements demonstrate that FL-P aqueous dispersions show strong visible and near-infrared emission that is characteristic of 1L, 2L, and 3L+ phosphorene. FL-P aqueous dispersions are also amenable to further size sorting by s-DGU, which allows for the enrichment of flakes with large lateral areas suitable for FET fabrication. The resulting FETs confirm that aqueous-exfoliated FL-P can be incorporated into electronic devices with performance metrics competitive with the best BP transistors to date. This demonstration of a scalable and environmentally friendly approach for isolating optically and electronically active phosphorene in aqueous solution will foster the development of phosphorene-based technologies in addition to suggesting a general methodology for exploring other chemically sensitive nanomaterials.

**Acknowledgements**

We gratefully acknowledge the assistance of D. Jariwala in measuring BP contact angles. Solution processing was supported by the National Science Foundation (DMR-1505849), structural and chemical characterization was supported by the Office of Naval Research (N00014-14-1-0669), and charge transport measurements were supported by the National Science Foundation Materials Research Science and Engineering Center (DMR-1121262). This work made use of the NUANCE Center, which has received support from the NSF MRSEC (DMR-1121262), the State of Illinois, and Northwestern University. The Raman instrumentation was funded by the Argonne-Northwestern Solar Energy Research (ANSER) Energy Frontier Research Center (DOE DE-SC0001059). The use of the near-infrared microscope at the Argonne National Laboratory Center for Nanoscale Materials was supported by the U. S. Department of Energy, Office of Science, Office of Basic Energy Sciences (Contract No. DE-AC02-06CH11357). S.A.W. was supported under contact FA9550-11-C-0028 from the Department of Defense, Air Force Office of Scientific Research, National Defense Science and Engineering Graduate (NDSEG) Fellowship. C.A.H. was supported by an Argonne National Laboratory named postdoctoral fellowship.

**Author Contributions**

J.K., S.A.W., and J.D.W. planned the experiments under the supervision of M.C.H. J.K. performed solution processing and characterization. S.A.W. prepared devices by EBL and measured charge transport. J.D.W. performed XPS. J.D.W. and J.K. collected PL and Raman spectra from FL-P dispersions. J.-H.L. and J.K. performed the AFM measurements and film transfer for FET fabrication. X.L. collected HRTEM and SAED data. C.R.R., J.D.W., and C.A.H. performed PL measurements on micromechanically exfoliated BP and solid-state BP with advice from J.R.G. J.Z. and J.K. collected zeta potential data. All authors contributed to the writing and critique of the manuscript.

**Figure Legends**

**Fig. 1**. Experimental procedure and characterization of FL-P nanosheets. (*A*) Schematic of the preparation method for FL-P aqueous dispersions. Deoxygenated water with 2% w v$^{-1}$ SDS was prepared by ultrahigh purity Ar purging. The BP crystal was exfoliated in a sealed container using tip ultrasonication, and then centrifuged to remove unexfoliated BP crystals. The FL-P dispersion was subsequently collected and ultracentrifuged to precipitate large flakes. The supernatant was finally redispersed in deoxygenated water. (*B*) Photographs of the as-exfoliated BP solution (left) and FL-P solution (right). (*C*) Low-magnification transmission electron microscopy (TEM) image of FL-P nanosheets. (*D*) High-resolution TEM image of a FL-P nanosheet. (*E*) Selected area electron diffraction (SAED) pattern of FL-P nanosheets. (*F*) Solution phase Raman spectrum of FL-P nanosheets. (*G*) Visible photoluminescence (PL) spectrum of FL-P nanosheets using an excitation wavelength of 532 nm with a Si CCD. (*H*) X-ray photoelectron spectroscopy (XPS) analysis of the FL-P nanosheets.

**Fig. 2**. Surface properties and exfoliation yield of BP in aqueous solution. (*A*) Photograph of a BP dispersion in deoxygenated water with and without SDS. (*B*) Photograph of a water contact angle measurement on an as-exfoliated flat BP crystal surface. (*C*) Optical absorbance spectra of BP dispersions with (red) and without (blue) SDS. (*D*) Zeta potential measurement of BP in water (blue), SDS-water (orange), and BP in SDS-water (red). (*E,F*) Photographs of BP dispersions in SDS-water and NMP after sonication and centrifugation at 0.5 k, 5 k, 10 k, and 15 k r.p.m. (*G*) Concentration of the BP dispersions from part (E) and (F). (*H,I*) AFM height images of BP nanosheets processed in SDS-water and NMP. (*J*) Thickness distribution of BP nanosheets in SDS-water (red) and NMP (green).

**Fig. 3**. Optical properties of FL-P aqueous dispersions. (*A*) Optical absorbance spectrum of FL-P and a higher resolution plot in the range between 900 nm and 1250 nm (inset). (*B*) PL spectra of FL-P aqueous dispersions. Three peaks corresponding to monolayer, bilayer, and three-plus layer phosphorene are observed at ~907 nm, ~1215 nm, and ~1428 nm,

respectively. (*C*) PL spectra of FL-P dispersions (red) and $O_2$-exposed FL-P dispersions (blue), and XPS spectra of $O_2$-exposed FL-P dispersions.

**Fig. 4.** Size sorting and electrical properties of FL-P aqueous dispersions. (*A*) Photographs of FL-P solution before (left) and after (right) size sorting using sedimentation-based density gradient ultracentrifugation (s-DGU). (*B*) Histogram of flake areas for fractions f1 to f6. (*C*) False-colored SEM image of an FL-P nanosheet FET with higher magnification image of the channel (inset). (*D*) Output curves for a FL-P FET. (*E*) Transfer curve for a FL-P FET plotted in linear scale (blue) and semi-log scale (green) for $V_{DS}$ values of -1 V (open circles) and -25 mV (closed circles). (*F*) Plot of $I_{ON}/I_{OFF}$ *versus* drive current that compares this work to previously reported BP FETs.

Figures

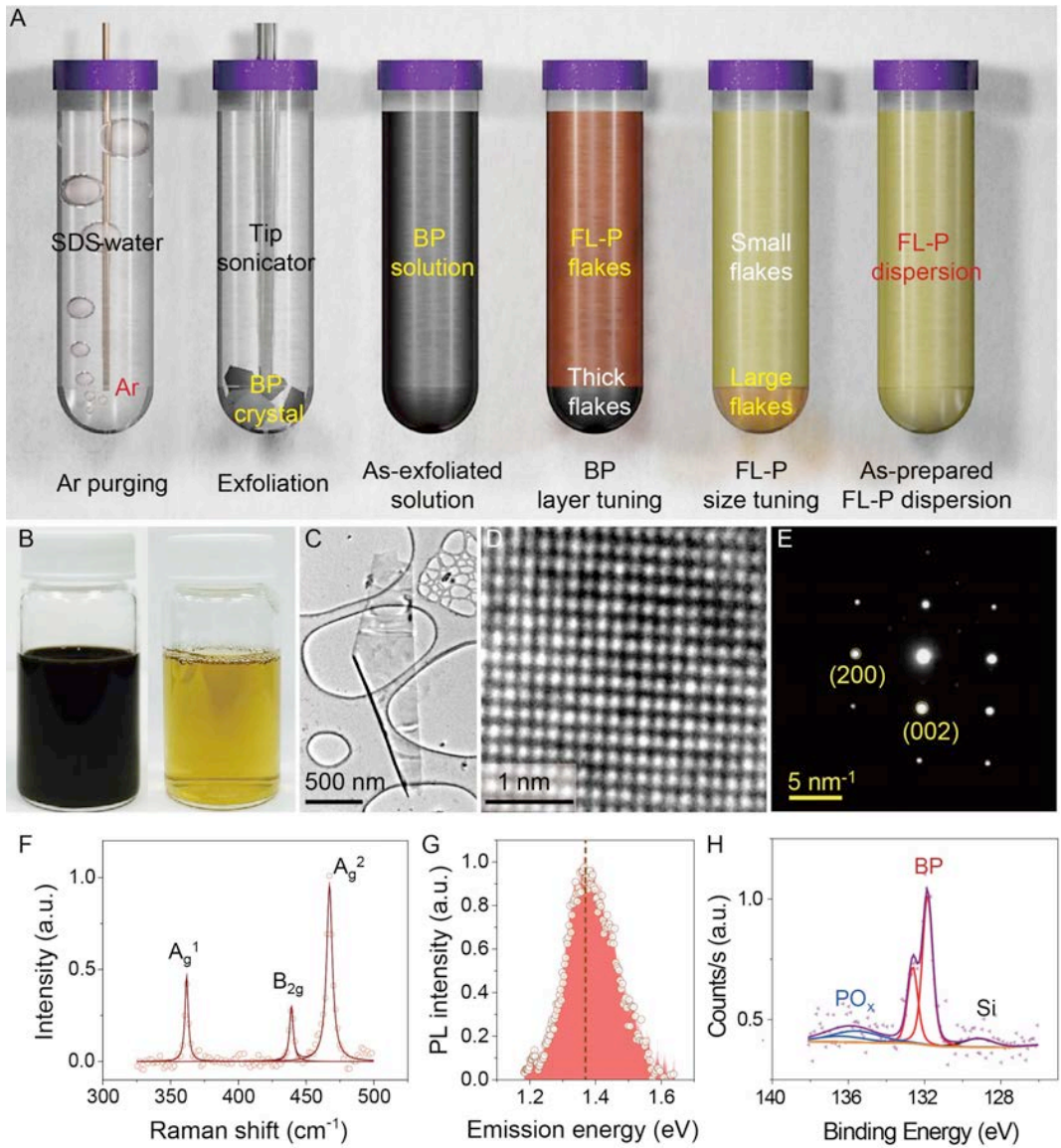

Figure 1.
Kang et al.

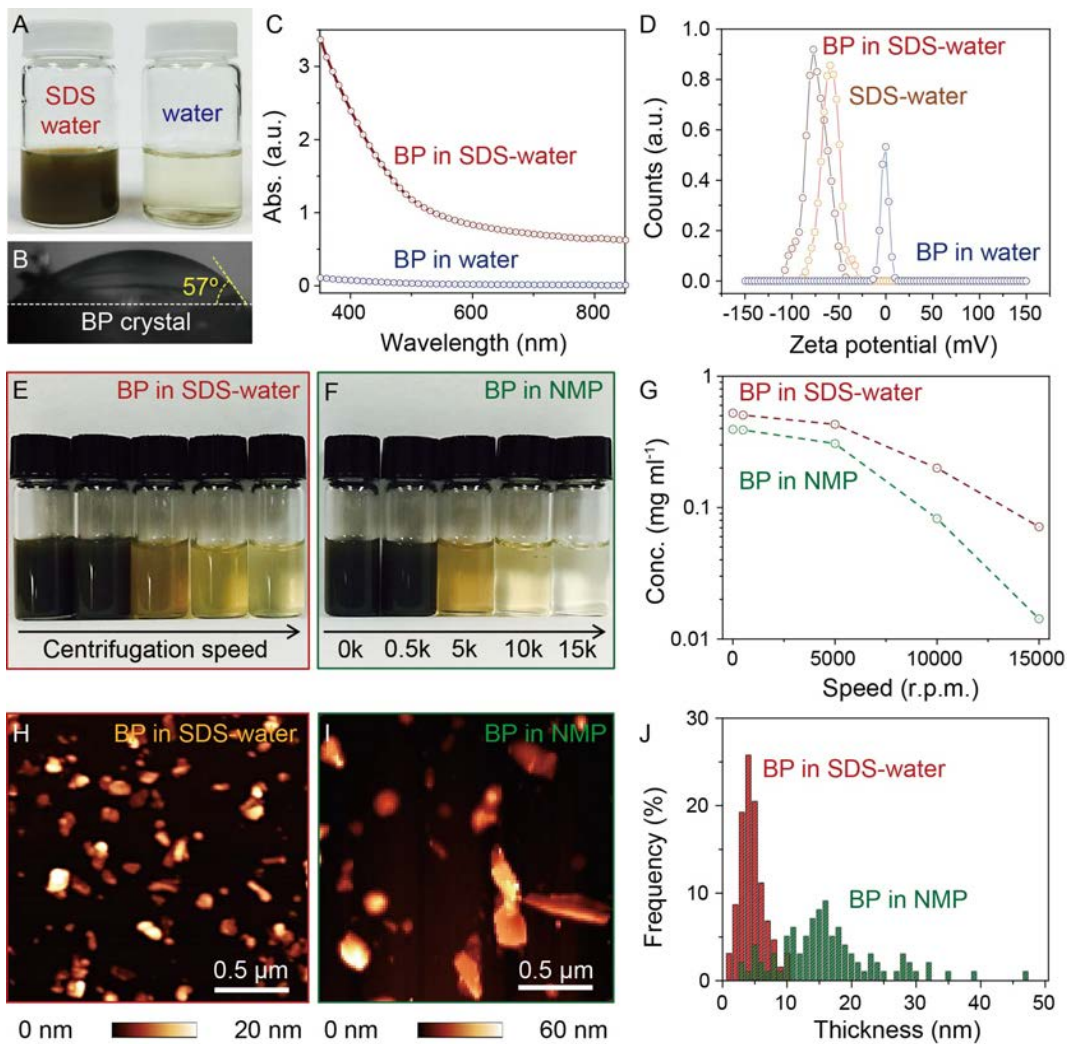

Figure 2. Kang et al.

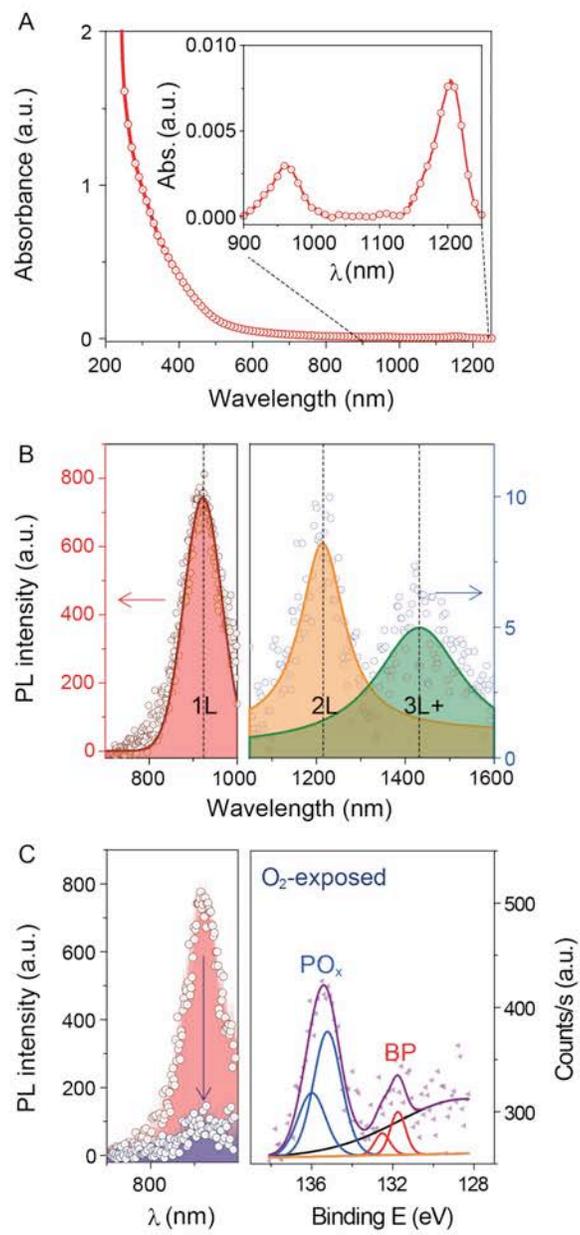

Figure 3.
Kang et al.

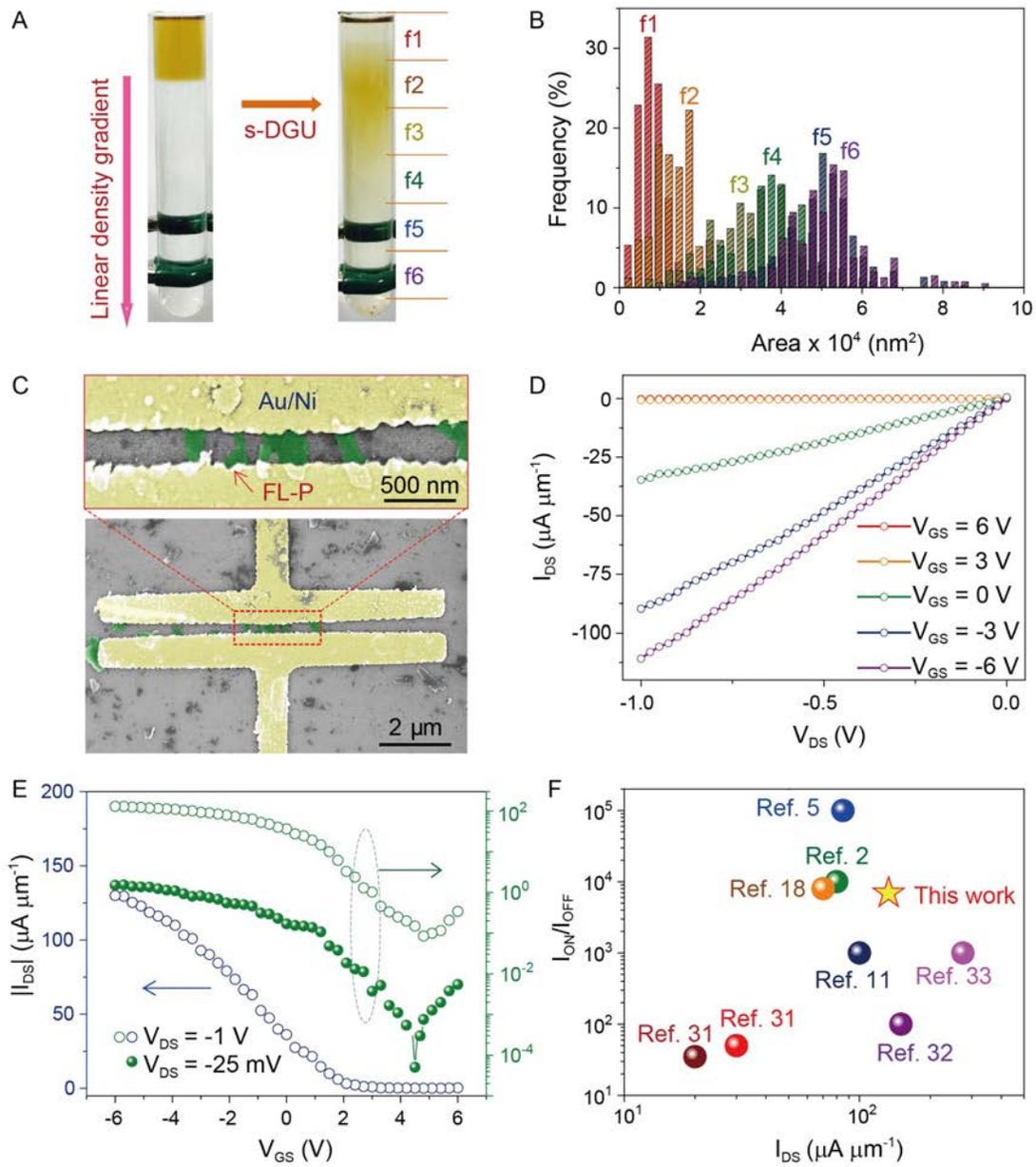

*Supplementary Information*

# Stable aqueous dispersions of optically and electronically active phosphorene

*Joohoon Kang[1], Spencer A. Wells[1], Joshua D. Wood[1], Jae-Hyeok Lee[1], Xiaolong Liu[2], Christopher R. Ryder[1], Jian Zhu[1], Jeffrey R. Guest[3], Chad A. Husko[3], and Mark C. Hersam[1,2,4,5,6]\**

[1]Department of Materials Science and Engineering, Northwestern University, Evanston, IL 60208, USA
[2]Graduate Program in Applied Physics, Northwestern University, Evanston, IL 60208, USA
[3]Center for Nanoscale Materials, Argonne National Laboratory, Argonne, IL 60439, USA
[4]Department of Chemistry, Northwestern University, Evanston, IL 60208, USA
[5]Department of Medicine, Northwestern University, Evanston, IL 60208, USA
[6]Department of Electrical Engineering and Computer Science, Northwestern University, Evanston, IL 60208, USA

\*Correspondence should be addressed to: m-hersam@northwestern.edu.

**Contents**
**Section S1. Methods**
- **S1.1. Few-layer phosphorene dispersion preparation**
- **S1.2. Contact angle measurement**
- **S1.3. Zeta potential measurement**
- **S1.4. Atomic force microscopy (AFM)**
- **S1.5. Transmission electron microscopy (TEM)**
- **S1.6. Raman spectroscopy**
- **S1.7. X-ray photoelectron spectroscopy (XPS)**
- **S1.8. Photoluminescence (PL) spectroscopy**
- **S1.9. Optical absorbance spectroscopy**
- **S1.10. FL-P nanosheets transfer by PDMS stamping**
- **S1.11. Field-effect transistor fabrication and measurement**

**Section S2. Supporting Figures and Data**
- **Fig. S1. Raman spectra of BP precipitants and few-layer phosphorene (FL-P)**
- **Fig. S2. XPS data for FL-P prepared with different surfactants**
- **Fig. S3. Extinction measurements of FL-P dispersions in sodium dodecylsulfate (SDS)-water**
- **Fig. S4. AFM derived flake thickness distributions**

- **Fig. S5. Photoluminescence (PL) spectra of exfoliated, monolayer phosphorene**
- **Fig. S6. PL spectra of bilayer phosphorene with three-layer and four-layer regions**
- **Fig. S7. Polarized PL spectra of three-layer and four-layer phosphorene**
- **Fig. S8. PL spectra of passivated, five-layer phosphorene**
- **Fig. S9. Solid-state PL spectra of BP nanosheets cast from solution**
- **Fig. S10. Infrared, solid-state PL spectra of solution cast BP nanosheets**
- **Fig. S11. Area differences for size-sorted FL-P nanosheets**
- **Fig. S12. AFM images of water rinsed BP nanosheets**
- **Fig. S13. XPS core level data for vacuum filtrated FL-P thin films**
- **Fig. S14. FET measurements of FL-P nanosheets**
- **Fig. S15. Electrical property histograms for FL-P FETs**
- **Table S1. Optical emissions for monolayer and few-layer phosphorene sheets**

## Section S1. Methods

### 1. Few-layer phosphorene dispersion preparation

Black phosphorus crystals were purchased from Smart-Elements and stored in an Ar glovebox with less than 1 ppm $O_2$. Deionized water with 2% w v$^{-1}$ surfactant was purged with ultrahigh purity grade Ar for at least 1 hr to remove dissolved oxygen. A customized tip sonicator setup was prepared by perforating the plastic lid of a 50 mL conical tube with a 0.125 inch sonicator tip. The interface between the tip and the lid was sealed with polydimethylsiloxane (PDMS) several times to occlude $O_2$ and $H_2O$. The prepared deoxygenated water and BP crystal were placed in this sealed conical tube with an initial concentration of 1 mg mL$^{-1}$ under an Ar atmosphere (<10% relative humidity). Additionally, Parafilm and Teflon tapes were used to further suppress ambient exposure. The sealed container was connected to the sonicator (Fisher Scientific model 500 sonic dismembrator) in ambient conditions. Subsequently, BP crystals were exfoliated by ultrasonication at 70 W for 1 h in an ice bath. Then, the solution was centrifuged at 7500 r.p.m. for 2 hr at 15 °C to enrich few-layer phosphorene (FL-P) nanosheets (Avanti J-26 XP, Beckman Coulter). Following centrifugation, supernatants were collected, and ultracentrifuged at 14,000 r.p.m. for 2 hr at 22 °C using a SW32Ti rotor (Optima L-80 XP, Beckman Coulter); these ultracentrifuged supernatants were finally redispersed in deoxygenated water.

## 2. Contact angle measurements

Water droplet contact angles on BP crystals were measured by a contact angle goniometer. Flat BP crystals were purchased from HQ Graphene (The Netherlands). Fresh BP surfaces were prepared by micromechanical exfoliation, after which water droplets were placed on the surface within ~30 s. A high-resolution camera attached on the goniometer captured images. From these images, contact angles were fitted with the low-bond axisymmetric drop shape analysis (LBADSA) technique in ImageJ (1).

## 3. Zeta potential measurement

Zeta potential measurements were carried out using a Zetasizer Nano ZS (Malvern Instruments) with clear disposable zeta cells. A He-Ne laser source with a wavelength of 633 nm and a maximum power of 5 mW was used for the measurements. The reported results are averages from three independent measurements at 25 °C.

## 4. Atomic force microscopy (AFM)

AFM images were acquired in tapping mode using an Asylum Cypher AFM with Si cantilevers (~290 kHz resonant frequency). As-prepared solutions were deposited onto Si substrates, rinsed with deoxygenated water to remove surfactant, and dried on a hot plate at 80 °C for 10 min in a flowing Ar environment. Prior to deposition, Si substrates were rinsed with acetone, methanol, and deionized water and immersed in diluted (3-aminopropyl)-triethoxysilane (APTES) solution to promote adhesion. After BP deposition on the Si substrate, an environmental cell was assembled in a flowing Ar environment and attached to a Cypher ES scanner under a laminar flow of ultrahigh purity grade $N_2$. Images were taken in the repulsive phase regime at a scanning rate of ~0.4 Hz using a minimum of 1024 samples per line. During scanning, $N_2$ was continuously flowed through the environmental cell under optical microscopy light illumination.

## 5. Transmission electron microscopy (TEM)

A BP solution droplet was deposited on a holey carbon TEM grid (Ted-Pella) and dried with $N_2$. The TEM grid was assembled with a TEM sample holder after fewer than 5 min of exposure to ambient air. The TEM images were taken with a JEOL JEM-2100 at an accelerating voltage of 200 keV with a TEM column pressure of ~$10^{-7}$ Torr.

## 6. Raman spectroscopy

Raman spectra of the solutions were obtained using a Horiba LabRAM HR Evolution with an excitation wavelength of 532 nm. A clear quartz cuvette with 10 mm transmitted path length was used for the measurement. Data were collected for 120 s at ~50 mW using an angled cuvette holder for the solution samples.

## 7. X-ray photoelectron spectroscopy (XPS)

XPS measurements were performed using a high vacuum Thermo Scientific ESCALAB 250 Xi XPS system at a base pressure of ~1×10$^{-9}$ Torr. The XPS data had a binding energy resolution of ~0.1 eV using a monochromated Al Kα X-ray source at ~1486.7 eV (~400 μm spot size). All core-level spectra were the average of five scans taken at a 100 ms dwell time using a pass energy of 15 eV. When using charge compensation, all core levels were charge-corrected to adventitious carbon at ~284.8 eV. Using the software suite Avantage (Thermo Scientific), all subpeaks were determined in a manner detailed elsewhere (2). The p core level spectra for phosphorus and silicon were fit with doublets.

## 8. Photoluminescence (PL) spectroscopy

PL spectra for the solution sample (Fig. 1G) were obtained simultaneously with Raman spectra (Fig. 1F) using a Horiba LabRAM HR Evolution with an excitation wavelength of 532 nm. Data were collected for 120 s at ~50 mW using an angled cuvette holder for the solution samples. PL spectra for the solution samples in Fig. 3B were obtained using a Horiba Fluorolog-3 spectrofluorometer. Data were measured in a quartz cuvette for 3 s. A liquid $N_2$ cooled InGaAs array was used for the spectra acquired at wavelengths longer than 1000 nm. PL spectra for the solid-state samples on 300 nm $SiO_2$/Si substrates were collected using a Horiba Xplora Raman/PL system with an excitation wavelength of 532 nm. Data were collected for 120 s using a 100× objective for point spectra and a 20× objective for mapping. This setup was used for the data in Figs. 3 and S5. For the infrared PL measurements in Figs. S6, S7, S8, and S10, a spectrometer (Princeton Aston) with a NIRvana InGaAs thermoelectrically cooled array at an excitation wavelength of 532 nm was employed. Data were collected for 5 s using a 100× near-infrared objective. All PL spectral fits were performed with spline baselines and Voigt (Gaussian-Lorentzian) functions. Spectral fits disregarded any lineshape changes due to differences in grating efficiency at off-blaze wavelengths. Estimating the dielectric environment around FL-P using $\varepsilon_{r,SDS}$ = 32 (3) and $\varepsilon_{r,DI}$ = 80 (4) for the dielectric constant of SDS and deionized water, respectively, results in the shifted PL emission energy.

### 9. Optical absorbance spectroscopy

Optical absorbance spectra were obtained using a Cary 5000 spectrophotometer (Agilent Technologies). A quartz cuvette with 10 mm path length was used for the measurements. The baseline from the aqueous surfactant solution was subtracted from the spectra.

### 10. FL-P nanosheets transfer by PDMS stamping

The few-layer phosphorene nanosheets were collected on anodic aluminum oxide (AAO) membranes with 100 nm pore size by vacuum filtration (Whatman™). Following vacuum filtration, the nanosheets on the membrane were rinsed with ~300 mL of deoxygenated water to remove the surfactants. To track the surfactant content before and after rinsing with deoxygenated water, AFM and XPS measurements were performed. In Fig. 12A, the BP nanosheet surface before deoxygenated water rinsing is covered with SDS (island-like features). In contrast, Fig. S12B shows a flat BP nanosheet surface after water rinsing. Fig. S13 shows that the BP P 2p doublet is maintained after rinsing, while the Na from the SDS is substantially reduced. Therefore, the post-filtration deoxygenated water rinsing significantly lowers the surfactant content. Finally, the nanosheets on the membrane were transferred to other recipient substrates by PDMS stamping.

### 11. Field-effect transistor fabrication and measurement

Field-effect transistors (FETs) were fabricated using electron beam lithography to define 200 nm long, 10 μm wide electrodes (5 nm Ni and 30 nm Au). BP FETs were measured in a Lakeshore CRX 4K under a base pressure of $<5\times10^{-4}$ Torr at room temperature. Two Keithley sourcemeter 2400 units were used to measure device performance. Equation 1 was used to measure carrier mobility:

$$\mu_{eff} = \frac{L g_d}{W C_{ox} V_{DS}} \quad (1)$$

where $\mu_{eff}$ is the field-effect mobility, $L$ is the channel length (obtained from optical micrographs), $g_d$ is the transconductance, $W$ is the channel width (obtained from optical micrographs), $C_{ox}$ is the oxide capacitance (measured values of $1.02 \times 10^{-2}$ and $2 \times 10^{-3}$ F·cm$^{-2}$ were used for 20 nm thick ALD HfO$_2$ and Al$_2$O$_3$, respectively), and $V_{DS}$ is the applied source-drain bias.

## Section S2. Supporting Figures and Data

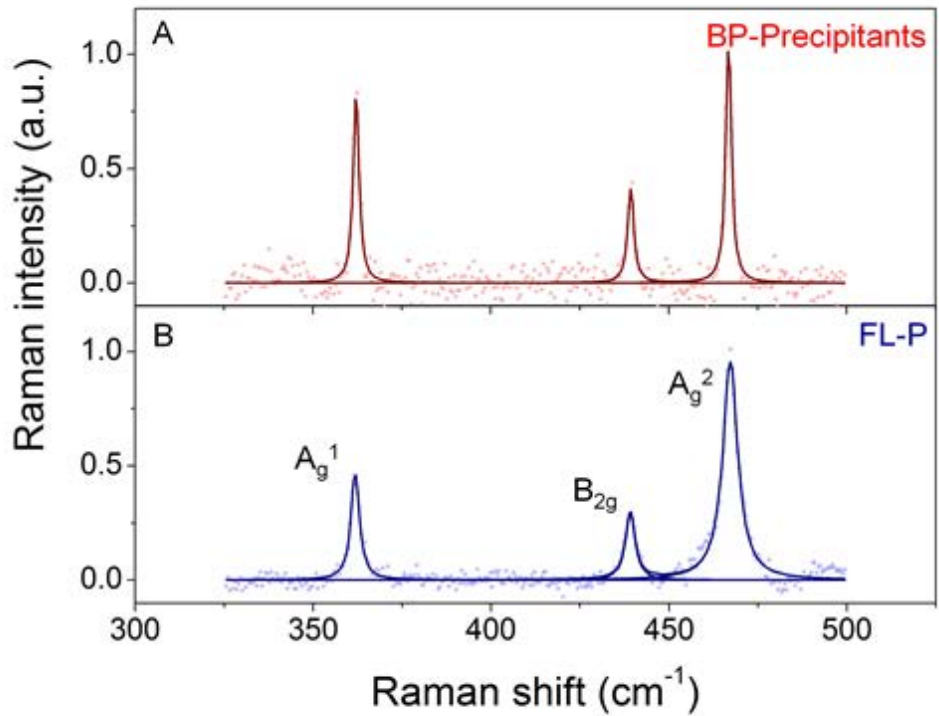

**Fig. S1.** Raman spectra of BP precipitants and few-layer phosphorene (FL-P). (*A*) The liquid-phase Raman spectrum from BP precipitants shows three modes at ~362 cm$^{-1}$, ~439 cm$^{-1}$, and ~467 cm$^{-1}$ with full width half maximum (FWHM) values of ~1.96 cm$^{-1}$, ~2.12 cm$^{-1}$, and ~1.95 cm$^{-1}$, respectively. (*B*) Conversely, the spectrum from FL-P shows three modes at ~362 cm$^{-1}$, ~439 cm$^{-1}$, and ~466 cm$^{-1}$ with FWHM values of ~2.96 cm$^{-1}$, ~3.31 cm$^{-1}$, and ~5.13 cm$^{-1}$, respectively.

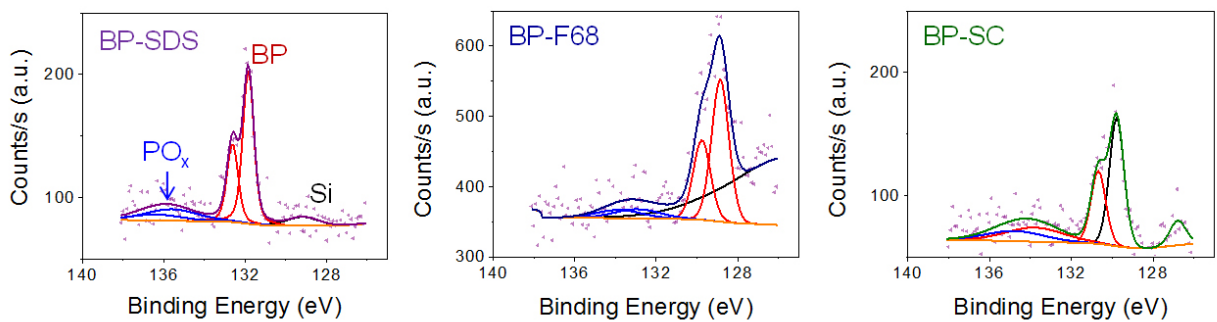

**Fig. S2.** XPS data for FL-P prepared with different surfactants. P 2p core level spectra for sodium dodecylsulfate (violet, left), Pluronic F68 (blue, middle), and sodium cholate (green, right).

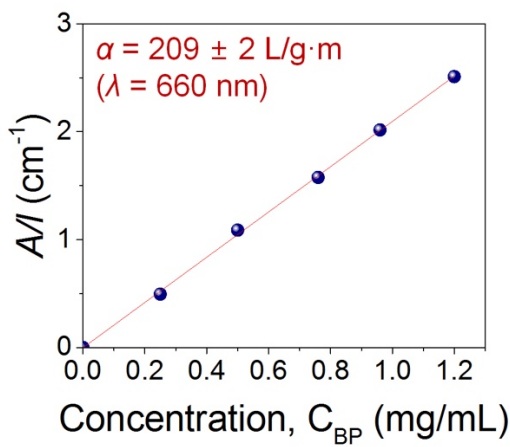

**Fig. S3.** Extinction measurements of FL-P dispersions in sodium dodecylsulfate (SDS)-water. The extinction coefficient is found to be 209 ± 2 Lg$^{-1}$m$^{-1}$, using the absorbance per length (*A/l*) at 660 nm *versus* the BP concentration. To calculate the concentration, FL-P dispersions with different volumes were filtered on an AAO membrane with 20 nm pore size, and then the mass difference was compared before and after filtration.

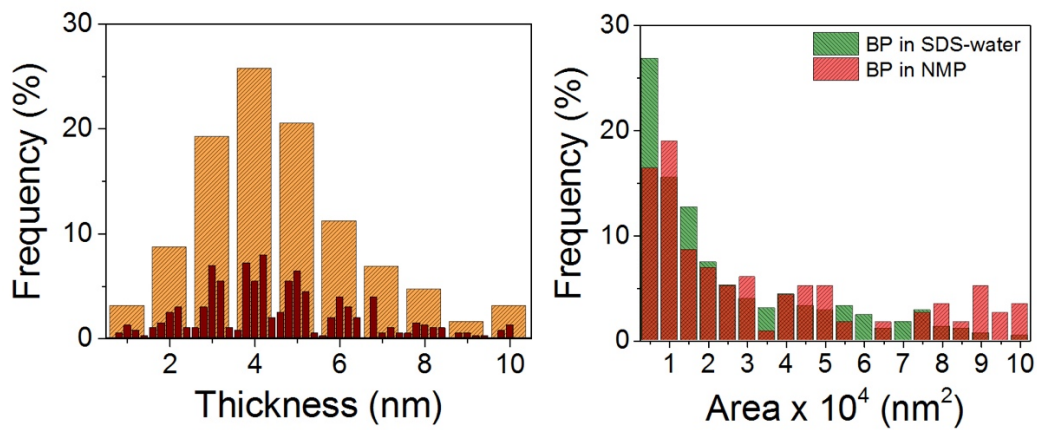

**Fig. S4.** AFM derived flake thickness distributions. (*left*) Thickness histogram and (*right*) area histogram of FL-P nanosheets prepared in SDS-water and in NMP (red). For the thickness histogram, binning is done with two layer thicknesses, employing ~1 nm layer thickness for the smaller scale (brown).

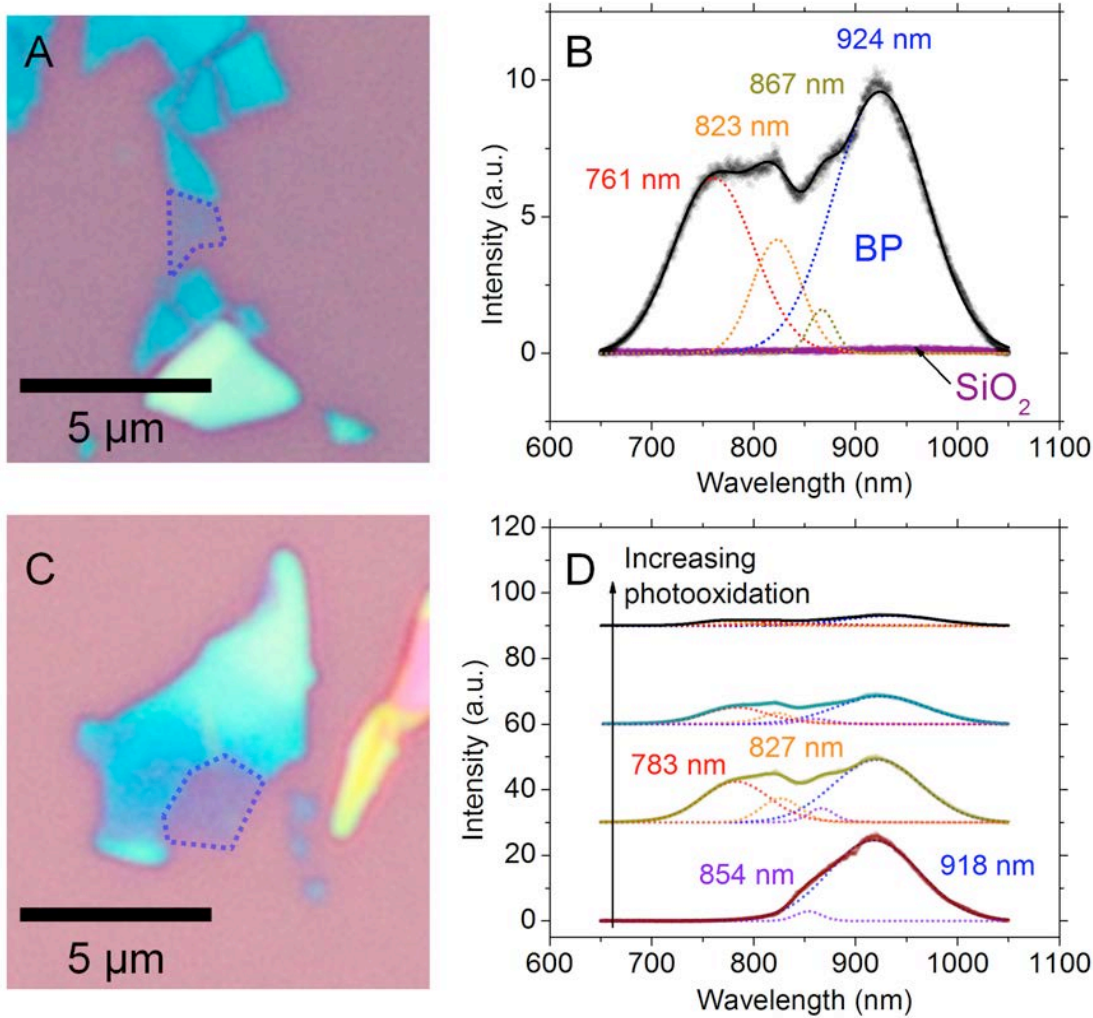

**Fig. S5.** Photoluminescence (PL) spectra of micromechanically exfoliated, monolayer phosphorene. (*A*) Optical image of a monolayer phosphorene flake (blue outline), passivated from ambient oxidation with ~2.6 nm of ~50 °C AlO$_x$ (5). (*B*) PL spectra of the phosphorene flake in (*A*), excited with a linearly polarized, 532 nm laser at ~15 µW (100×, 0.9 NA objective, 10 s acquisition in air). The phosphorene monolayer emits at 924 nm (~1.34 eV), near the expected wavelength (6), whereas emissions from the SiO$_2$ substrate (purple) are absent. Additional emissions in the visible are present (761 nm, 823 nm, and 867 nm, respectively), potentially related to oxidized BP regions (7). (*C*) Optical image of a second phosphorene region (blue outline), passivated like (a). (*D*) PL spectra of the flake in (*C*) after successive spectral acquisitions with a 532 nm laser at ~15 µW (100×, 0.9 NA objective, 20 s acquisition in air). While the phosphorene monolayer has its excitonic emission at ~918 nm (~1.35 eV), it possesses a small oxide peak at ~854 nm (~1.45 eV). With additional measurements, the monolayer phosphorene emission decreases in intensity with the concurrent emergence of oxide-related bands (ca. 783 nm and 827 nm, respectively), indicative of photooxidation (8).

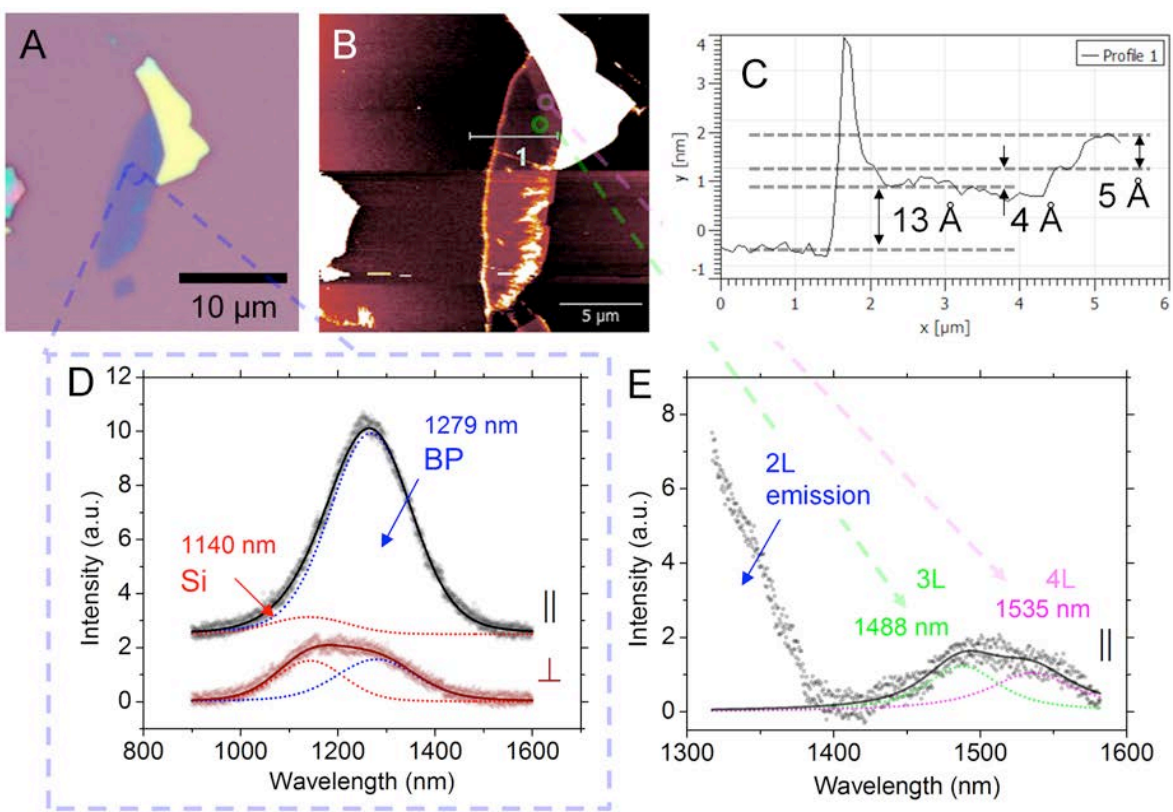

**Fig. S6.** PL spectra of micromechanically exfoliated bilayer phosphorene with three-layer and four-layer regions. (*A*) Optical image of a bilayer phosphorene flake (dark blue) attached to a thick BP region, passivated by ~10.0 nm AlO$_x$ (~2.5 nm at 50 °C ALD, ~7.5 nm at 150 °C ALD).(5) (*B*) AFM height image of the flake in (*A*), showing bilayer, three-layer, and four-layer phosphorene areas. (*C*) Height profile along the contour in (*B*). Bilayer (~1.3 nm), three-layer (~1.7 nm), four-layer (~2.2 nm) steps indicated (~0.5 nm from instrumental factors, AlO$_x$ overlayer homogeneity, and adhesion). (*D*) PL spectra of the bilayer phosphorene (blue spot, *A*), excited with a 532 nm laser at ~150 μW (50×, 0.8 NA objective, 5 s acquisition in air). The laser polarization was adjusted to maximize the bilayer emission relative to the ~1143 nm (~1.08 eV) Si emission (termed "parallel" orientation: ∥). Under parallel orientation, the bilayer phosphorene emits at 1279 nm (~0.97 eV). To confirm the anisotropic optical properties of BP, the bilayer emission was dampened five-fold by rotating the laser polarization 90° (±20°) relative to the parallel orientation (termed "perpendicular" orientation: ⊥). (*E*) PL spectrum from three-layer (green spot) and four-layer (magenta spot) phosphorene steps of (*B*), excited like (*D*). Spectrum taken from a single area on the bilayer phosphorene. While the bilayer phosphorene emission persists, emissions occur at 1488 nm (~0.83 eV) and 1535 nm (~0.81 eV) for three-layer and four-layer phosphorene, respectively.

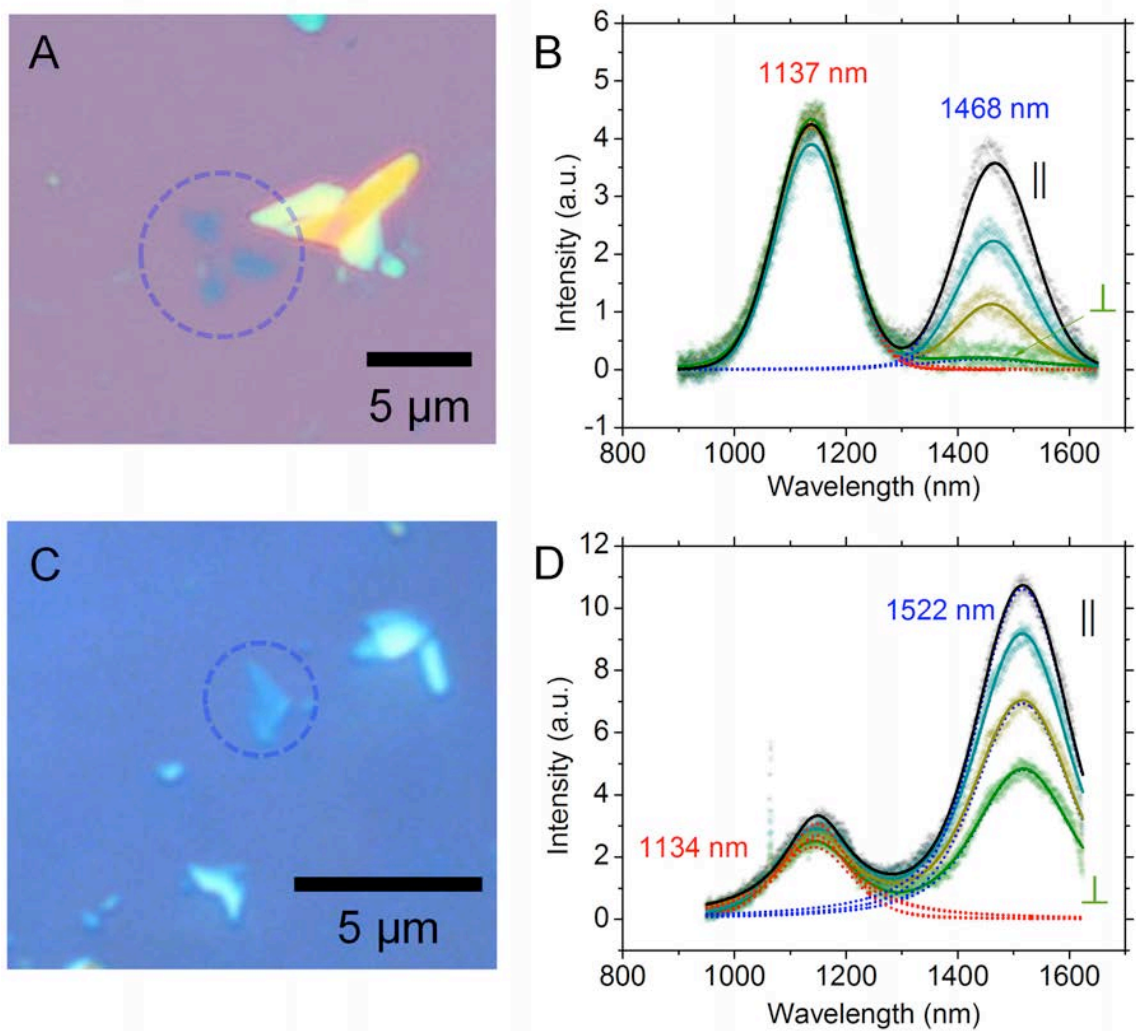

**Fig. S7**. Polarized PL spectra of micromechanically exfoliated three-layer and four-layer phosphorene. (*A*) Optical image of a three-layer phosphorene flake, passivated by ~10.0 nm AlO$_x$ (~2.5 nm at 50 °C ALD, ~7.5 nm at 150 °C ALD) (5). (*B*) Polarization dependent PL spectra for the three-layer phosphorene in (*A*), ranging from minimum ("perpendicular") to maximum ("parallel") phosphorene emission (blue-dotted) at ~1468 nm (~0.84 eV). Polarization invariant Si emission (red-dotted) at ~1137 nm (~1.09 eV). The sample was excited with a 532 nm laser at ~200 µW (50×, 0.8 NA objective, 5 s acquisition in air). (*C*) Optical image of a four-layer phosphorene flake, encapsulated like (*A*). (*D*) Polarization dependent PL spectra for the four-layer phosphorene in (*C*), with maximum phosphorene emission (blue-dotted, "parallel") at ~1522 nm (~0.81 eV) and Si emission (red-dotted) at ~1134 nm (~1.09 eV). The sample was excited at ~250 µW with all other exposure conditions being the same as (*B*).

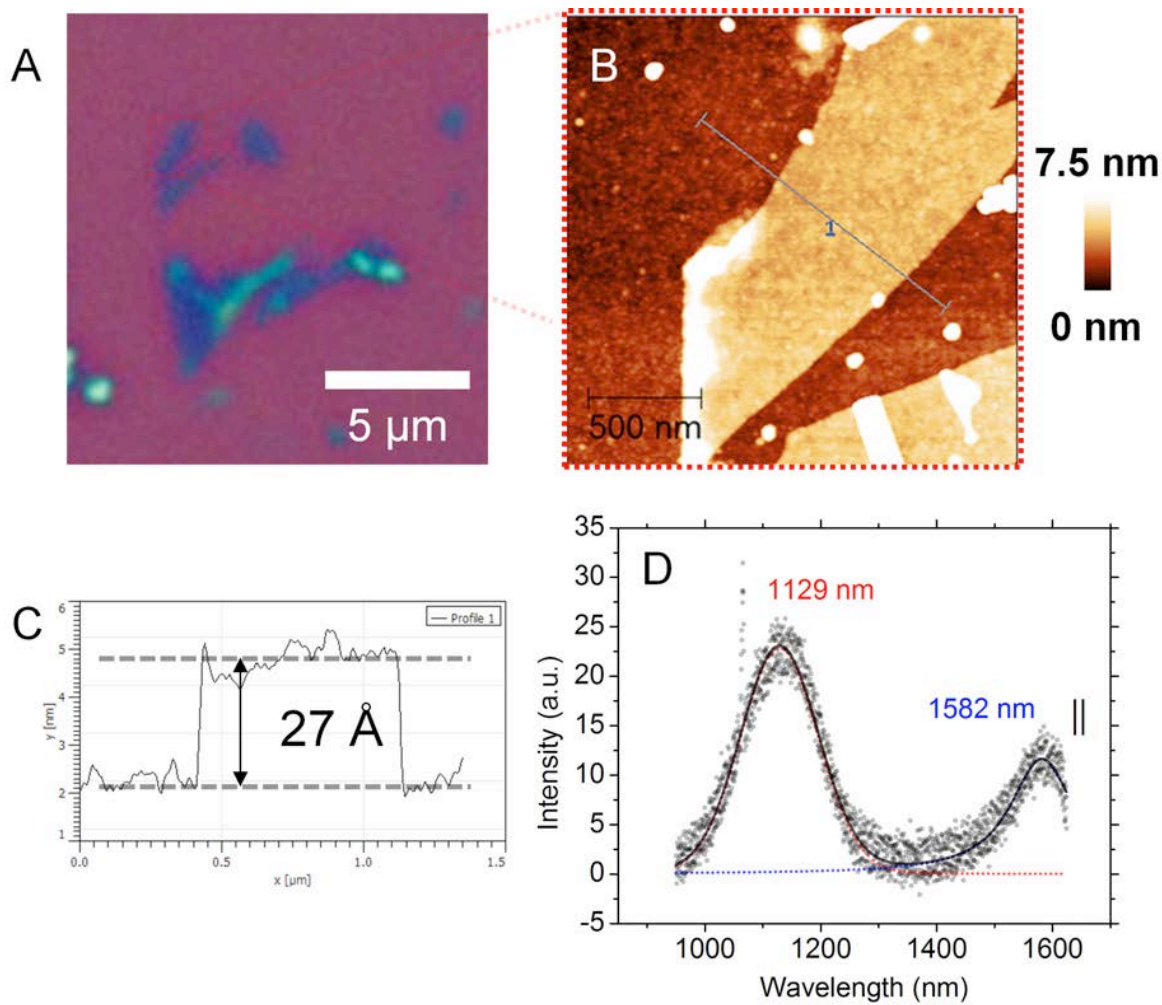

**Fig. S8.** PL spectra of micromechanically exfoliated, passivated, five-layer phosphorene. (*A*) Optical image of a five-layer phosphorene flake, passivated by ~12.6 nm AlO$_x$ (~2.6 nm at 50 °C ALD, ~10 nm at 150 °C ALD)(5). (*B*) AFM height image of the boxed region in (*A*). (*C*) Height profile along the line in (*B*). Passivated phosphorene flake is ~2.7 nm tall, indicative of five-layer phosphorene (~0.5 nm additional, as aforementioned). (*D*) PL spectrum of the phosphorene flake in (*B*), excited in the "parallel" orientation with a 532 nm laser at ~150 μW (50×, 0.8 NA objective, 5 s acquisition in air). Passivated, five-layer phosphorene emits (blue) at 1582 nm (~0.78 eV), and the Si substrate emits (red) at 1129 nm (~1.10 eV), as expected.

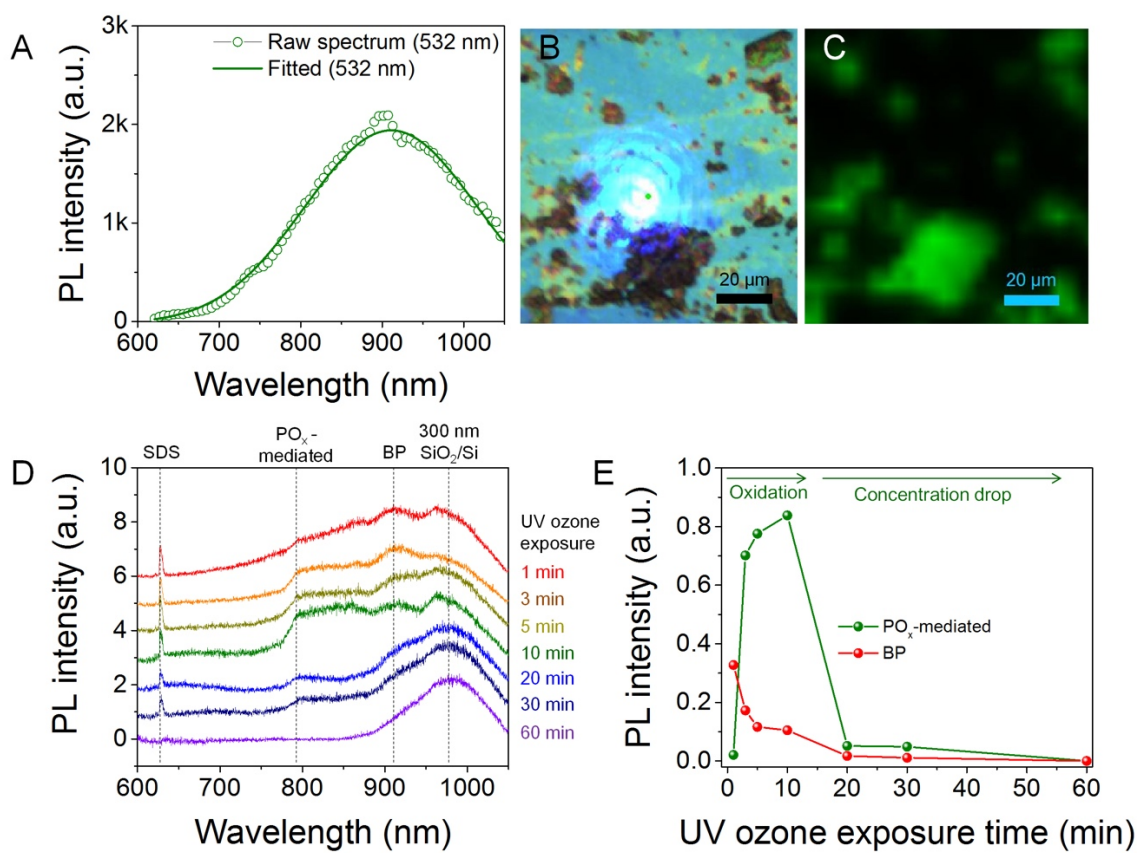

**Fig. S9**. Solid-state PL spectra of BP nanosheets cast from solution. (*A*) PL spectrum of FL-P aggregates on a 300 nm $SiO_2$/Si substrate. (*B*) Optical image of FL-P aggregates on the substrate. (*C*) Overlaying (*B*) with a (*C*) ~909 nm PL peak intensity map indicates that the PL emission comes from the aggregated FL-P. (*D*) PL spectra with respect to UV ozone exposure time. With increasing exposure, the PL emission intensity at ~909 nm decreases from monolayer phosphorene oxidation. Furthermore, an emission peak develops at ~780 nm (~1.59 eV), likely related to $PO_x$ species (4). Finally, a SDS Raman band disappears with 60 min UV exposure. (*E*) Emission intensities at ~780 nm (green) and ~909 nm (red) as a function of UV ozone exposure time. The peak at ~909 nm decreases monotonically as the FL-P is oxidized. The peak at ~780 nm increases initially but ultimately decays from extended chemical modification.

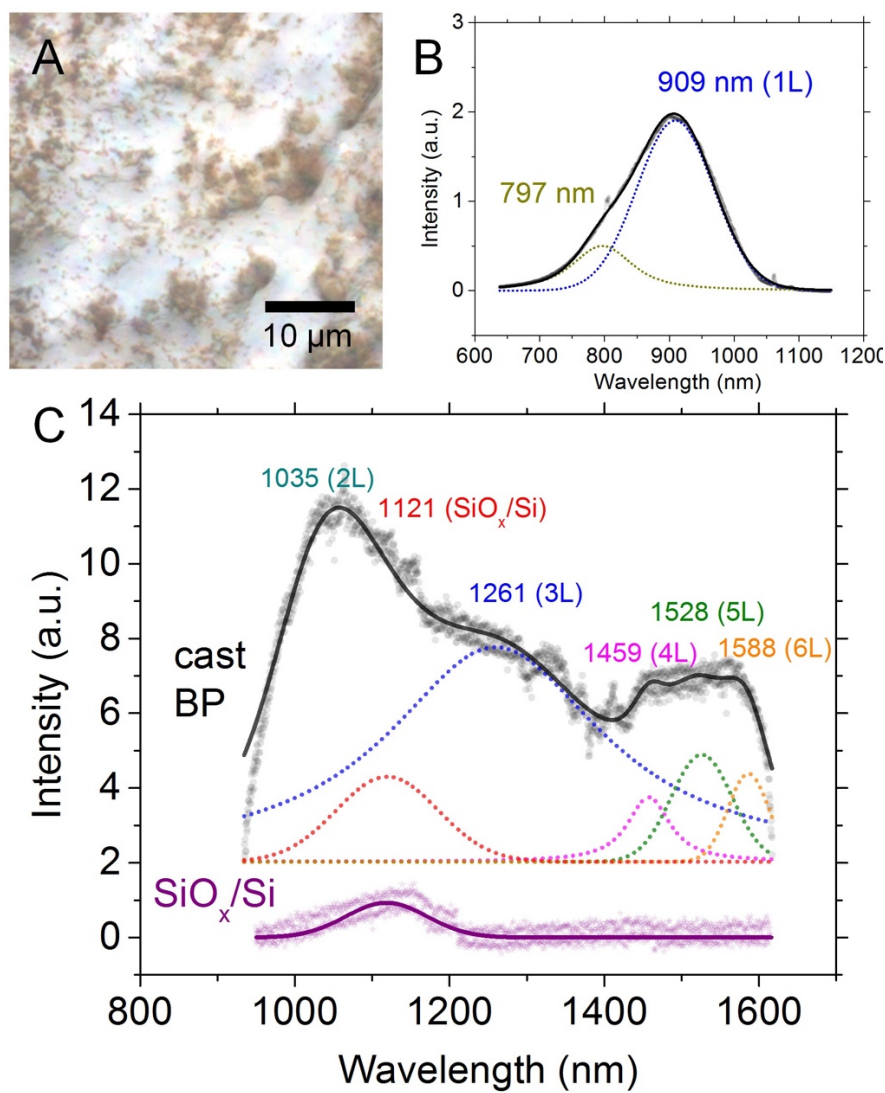

**Fig. S10.** Infrared, solid-state PL spectra of solution cast BP nanosheets. (*A*) Optical image of unencapsulated FL-P aggregates on native oxide (SiO$_x$) on Si(100). (*B*) Visible PL spectrum for the aggregates in (*A*), exposed to a ~350 µW, 532 nm laser (50×, 0.8 NA objective, 15 s acquisition in air). Monolayer phosphorene emission is present at 909 nm (~1.36 eV), with an additional emission at ~797 nm (~1.56 eV) (7). Spectrum agrees well with the FL-P aggregate emissions of Fig. S9. (*C*) Infrared PL spectra for the aggregates in (*A*). Control SiO$_x$/Si(100) spectrum (purple) shows a weak Si emission (red) at ~1121 nm (~1.11 eV). FL-P aggregates are broadband emitters, having bilayer (1261 nm, ~0.98 eV), three-layer (1459 nm, ~0.85 eV), four-layer (1528 nm, ~0.81 eV), and five-layer (1588 nm, ~0.78 eV) phosphorene emissions. A band tail from monolayer phosphorene (see (*B*)) also exists (peak at 1035 nm given for reference). Solution-cast phosphorene emission energies are consistent with the mechanically exfoliated BP spectra of Figs. S5-S8. All peak emission wavelengths determined by Voigt fits to the baseline-corrected spectra.

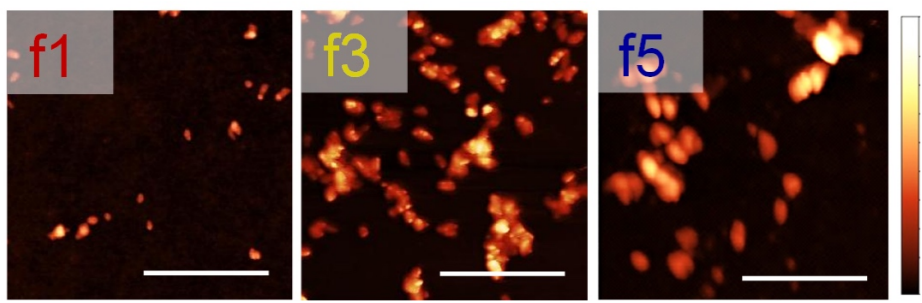

**Fig. S11.** Area differences for size-sorted FL-P nanosheets. AFM images of fractions 1, 3, and 5 after sedimentation-based density gradient ultracentrifugation (scale bars: 1 µm; height: 10, 30, and 30 nm, respectively).

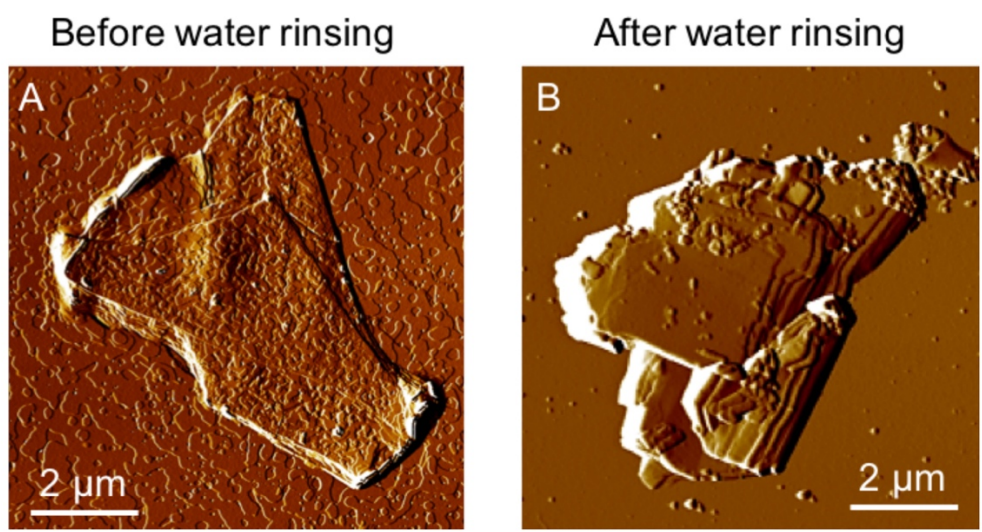

**Fig. S12.** AFM images of water rinsed BP nanosheets. While the surface of flake (*A*) before deoxygenated water rinsing is covered with SDS (island-like features), the surface is flat (*B*) after water rinsing, suggesting removal of SDS.

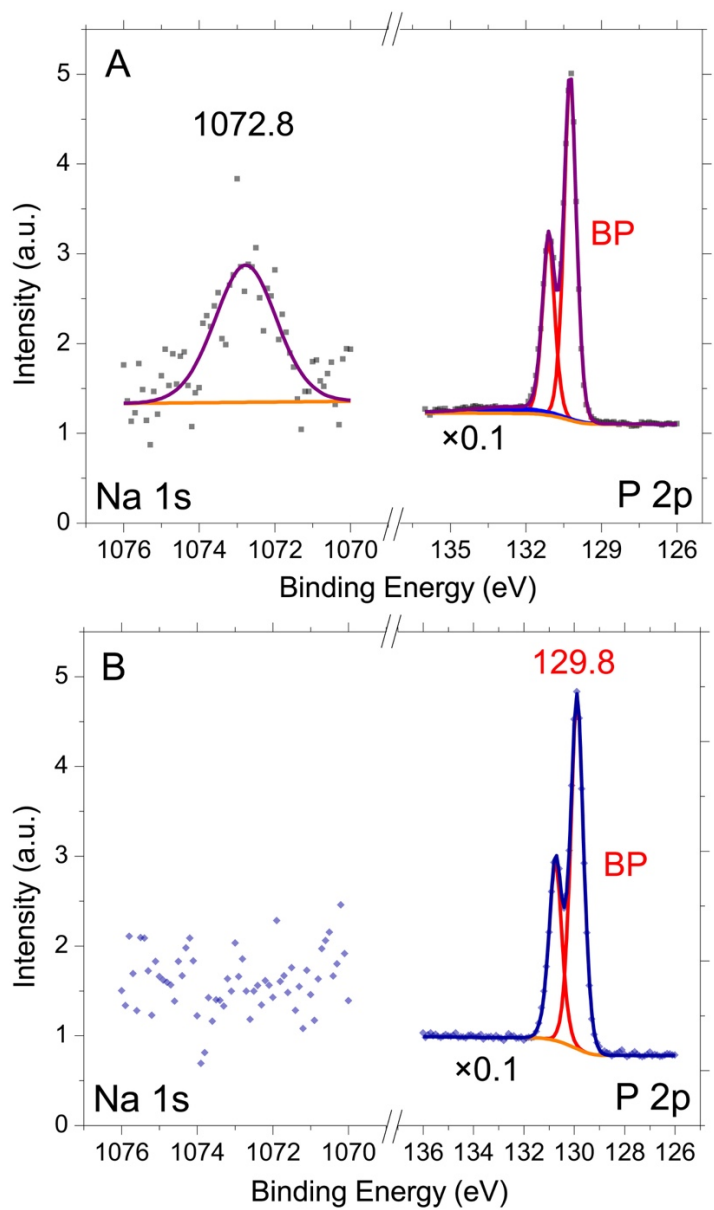

**Fig. S13.** XPS core level data for vacuum filtrated FL-P thin films. Na 1s and P 2p core level spectra for FL-P thin films (*A*) before sodium dodecyl sulfate (SDS) rinsing and (*B*) after SDS rinsing with deoxygenated water. The BP P 2p doublet is maintained after deoxygenated water rinsing, while the Na from the SDS is reduced substantially.

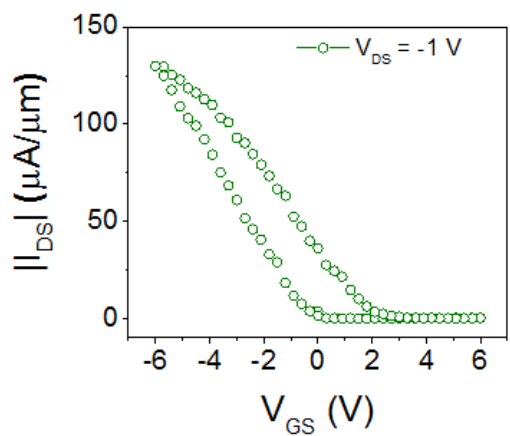

**Fig. S14.** FET measurements of FL-P nanosheets. This FET transfer curve shows the forward and reverse sweep for the device in Fig. 4*E*.

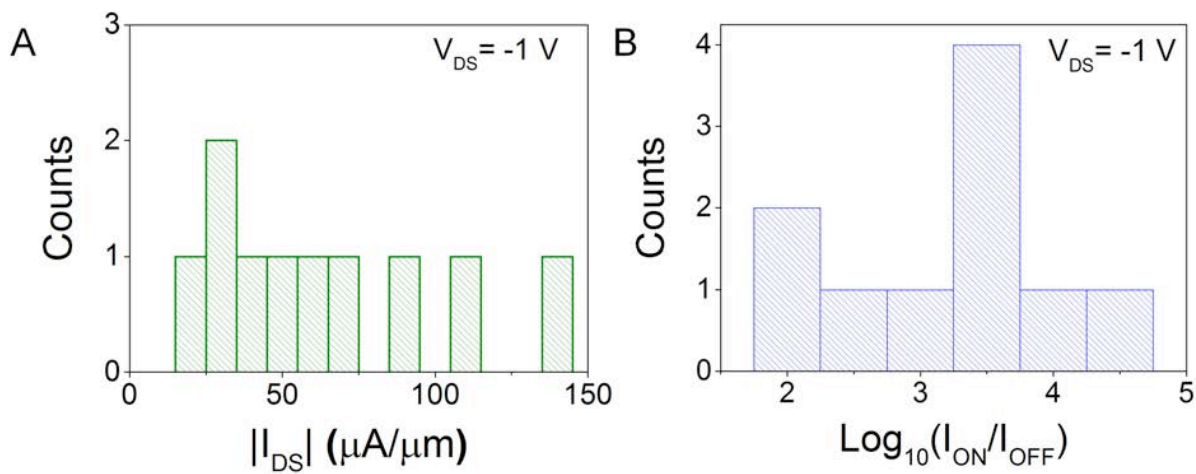

**Fig. S15.** Electrical property histograms for FL-P FETs. Histogram of (*A*) drive current, and (*B*) $I_{ON}/I_{OFF}$ ratio with logarithmic scale.

| Layer number | Isolation scheme | Primary emission (eV) |
|---|---|---|
| Monolayer | Mechanical exfoliation | 1.34 ± 0.15 eV |
| Bilayer | Mechanical exfoliation | 0.97 ± 0.16 eV |
| Three-layer | Mechanical exfoliation | 0.85 ± 0.09 eV |
| Four-layer | Mechanical exfoliation | 0.82 ± 0.10 eV |
| Five-layer | Mechanical exfoliation | 0.78 ± 0.06 eV |
| Monolayer | Solution cast | 1.38 ± 0.18 eV |
| Bilayer | Solution cast | 0.98 ± 0.29 eV |
| Three-layer | Solution cast | 0.85 ± 0.03 eV |
| Four-layer | Solution cast | 0.81 ± 0.05 eV |
| Five-layer | Solution cast | 0.78 ± 0.03 eV |

**Table S1.** Optical emissions for monolayer and few-layer phosphorene sheets. All emissions probed by solid-state PL spectroscopy. Errors are propagated from the FWHM values of each phosphorene emission.